%% file: ms.tex
\documentclass[runningheads]{llncs}

\usepackage[utf8]{inputenc}
\usepackage{url}
\usepackage{graphicx}
\usepackage{booktabs}

\usepackage{amsmath}
\usepackage{breakurl}
\usepackage{hyperref}
\usepackage[lofdepth,lotdepth,caption=false]{subfig}
\usepackage[]{todonotes}
\usepackage{comment}
\usepackage{xspace}

\usepackage{makecell}  

\begin{document}

  \author{Matteo Romiti\inst{1}\and
  Aljosha Judmayer\inst{2} \and
  Alexei Zamyatin\inst{2}\textsuperscript{,} \inst{3} \and
  Bernhard Haslhofer\inst{1}}
  \authorrunning{Romiti et al.}
  %
  \institute{Austrian Institute of Technology \email{\{name.surname\}@ait.ac.at} \and
  SBA Research \email{ajudmayer@sba-research.org} \and Imperial College London
  \email{a.zamyatin@imperial.ac.uk}}

  \title{A Deep Dive into Bitcoin Mining Pools}
  \subtitle{An Empirical Analysis of Mining Shares}

  \maketitle

\begin{abstract}
Miners play a key role in cryptocurrencies such as Bitcoin: they invest substantial computational resources in processing transactions and minting new currency units.
It is well known that an attacker controlling more than half of the network's mining power could manipulate the state of the system at will. 
While the influence of large mining pools appears evenly split, the actual distribution of mining power within these pools and their economic relationships with other actors remain undisclosed. 
To this end, we conduct the first in-depth analysis of mining reward distribution \emph{within} three of the four largest Bitcoin mining pools and examine their cross-pool economic relationships.
Our results suggest that individual miners are simultaneously operating across all three pools and that in each analyzed pool a small number of actors ($ \leq 20 $) receives over $ 50 \% $ of all BTC payouts.
While the extent of an operator's control over the resources of a mining pool remains an open debate, our findings are in line with previous research, pointing out centralization tendencies in large mining pools and cryptocurrencies in general.

\end{abstract}

  \def\currentBlockheight{556400}
  \def\percentMultiCBOut{5.1} 
  \def\srcBlockchainInfo{\url{https://raw.githubusercontent.com/blockchain/Blockchain-Known-Pools/29ab27c844ebdb63110f8783f73b9decd4abc221/pools.json}} 
  \def\srcBtcCom{\url{https://raw.githubusercontent.com/btccom/Blockchain-Known-Pools/650a92227bf65b06ff0a5b58bb57c13856a3babf/pools.json}} 
  \def\numberConflicts{684} 
  \def\percentConflicts{0.0012} 
  \def\attrStart{500000} 
  \def\attrEnd{556400} 
  \def\attr{54409} 
  \def\attrMoreBlockchainInfo{2568} 
  \def\attrMoreBlockchainInfoBTC{32100} 
  \def\stackplotStart{2013-12-21} 
  \def\stackplotEnd{2018-12-19} 

\input{sections/introduction}

\input{sections/background}
\input{sections/analysis}

\input{sections/discussion}

\input{sections/conclusions}

\input{sections/acknowledgments}



\input{ms.bbl}


\end{document}

%% file: sections/introduction.tex
\section{Introduction}\label{sec:introduction}


The distribution of mining power or \emph{hash rate} can be seen as a  key indicator for the market shares of mining pools and represents a core security parameter in Bitcoin and other cryptocurrencies relying on Nakamoto consensus
~\cite{nakamoto2008bitcoin,garay2016bitcoin,rosenfeld2014doublespending,sompolinsky2016bitcoin}.
An attacker in control of the majority of the network's hash rate is capable of manipulating the system at will, i.e., executing double spending attacks, prohibiting transactions from entering the blockchain and effectively rewriting the transaction history within computational limits.
However, concentrations even below this barrier open up the possibility for effective selfish-mining attacks and its variants~\cite{eyal2014majority,nayak2015stubborn,sapirshtein2015optimal,gervais2016security}. 
Since the hash rate distribution is not directly observable in the public blockchain, it is currently estimated retroactively by attributing mined blocks to known mining pools and counting their relative frequency. 

Statistics published by popular analytics platforms\footnote{Including but not limited to~\url{https://blockchain.info/charts}} indicate that the overall mining power is concentrated among a relatively small number of pools, with BTC.com, ViaBTC and AntPool holding―or being close to hold―the majority, but none of them exceeding the 50\% limit. 
However, large miners and pools could have an incentive to conceal or obfuscate the actual extent of their mining power. 
If successful, this would allow to maximize market shares and profits without visibly harming the security and credibility of the underlying system, ideally maintaining the stability of revenue streams. 


Despite its relevance for security considerations, measurement studies on the level of mining-power centralization are still scarce.
Recently, a study of miner centralization on the P2P network layer by Gencer et. al~\cite{gencer2018decentralization} found that over 50\% of the mining power has exclusively been shared by eight miners in Bitcoin and five miners in Ethereum for prolonged periods. 
In~\cite{ozisik2017estimation}, two novel methods to accurately estimate network hash rates are proposed, however, without providing empirical results. 
Judmayer et al. perform an analysis of mining power distribution in merge-mined cryptocurrencies, showing the latter exhibits vulnerabilities to centralization by large mining pools~\cite{judmayer2017merged}.
A series of blog posts on the distribution of Bitcoin mining shares covering the period between early 2012 and late 2016~\cite{online:organofcorti} are consistent with the results reported by Gencer et. al~\cite{gencer2018decentralization}.\newline


\noindent\textbf{Contributions.}
In this paper we expand on existing research, aiming at shedding light on the distribution of mining rewards in Bitcoin.
Specifically, we conduct an empirical analysis of mined blocks and mining reward payout patterns to present a clearer picture on how mining pools operate and how the mining reward assigned to operators is further distributed to individual miners enabling the analysis of monetary flows and economic relationships.
We leverage attribution techniques described in~\cite{judmayer2017merged}, improve their accuracy by incorporating information from existing online analysis platforms~\cite{online:blockchaininfo,online:blocktrail} and apply multiple-input clustering~\cite{haslhofer2016bitcoin,github:graphsense} for identifying address clusters that are likely to be controlled by the same actor.
Summarizing, our contributions are as follows:
\begin{enumerate}

  	\item We combined several sources of miner and pool attribution data and present a simple yet effective method for attributing blocks to mining pools. Using this method, we were able to attribute approximately 35K blocks more than \emph{blockchain.info} and analyzed the longitudinal evolution of the mining pool market share distribution in Bitcoin over a 5-year period. 

	\item We investigate payout patterns of three mining pools that together exceeded the 50\% threshold (between the end of 2017 and mid-2018) and analyze how mining rewards are distributed from pools to individual miners within a four-week observation period. Our results show that in each of the three pools, a small number of address clusters ($\leq 20$) receives over 50\% of all payouts, suggesting a relatively strong centralization.

	\item We examine economic relationships of miners, reveal cross-pool mining activities and find that geographically co-located exchange services and wallet providers are major receivers of minted coins. 

	\item We publicly release our findings on addresses, pools and mining entities, as well as the code to reproduce our results and improve the current knowledge about Bitcoin mining and involved actors at \url{https://github.com/MatteoRomiti/Deep_Dive_BTC_Mining_Pools}.

\end{enumerate}

\noindent\textbf{Outline.}
First, in Section~\ref{sec:background}, we introduce the necessary background on Bitcoin mining and mining pools. Then we go on and present our empirical analysis and the results we obtained in Section~\ref{sec:analysis}. We finish by discussing our findings in Section~\ref{sec:discussion} and conclude in Section~\ref{sec:conclusions}.

%% file: sections/background.tex
\section{Background}\label{sec:background}

In this section, we briefly introduce central notions used throughout this paper. While we do not attempt to give a complete introduction to the underlying technology of Bitcoin and permissionless cryptocurrencies, we direct the reader to existing literature, such as~\cite{nakamoto2008bitcoin,bonneau2015research,tschorsch2015bitcoin,narayanan2016bitcoin}.

\subsection{Bitcoin Mining and Mining Pools}

One of the key innovations of Bitcoin's Nakamoto consensus is the successful implementation of a random leader election process in a dynamically changing set of pseudonymous participants. Thereby, each node taking part in the consensus mechanism is required to solve a memoryless and non-invertible cryptographic puzzle, i.e., provide a Proof-of-Work (PoW). In Bitcoin, the latter is represented by a partial-preimage attack on the SHA-256 algorithm, whereby participants must brute-force the hash
over the transactions defining the new state of the system and some additional information, including the reference to the last seen block and a random nonce. To become leader, participants, also referred to as \emph{miners}, must provide a hash which is not only valid but also fulfills the network's current difficulty requirements, i.e., lies below a specified target value. 
The difficulty is adjusted every 2016 blocks (est. two weeks), such that the average interval at which PoW solutions are found is approximately 10 minutes.  

Due to the structure of PoW, the time between each two found blocks in Bitcoin is exponentially distributed, yielding the number of found blocks per time period a Poisson process. The rate parameter is thereby defined by the ratio of PoW difficulty to the overall mining power present in the network. As such, \emph{solo miners} are expected to face a high variance of payouts, depending on their share of the overall mining power. Consequently, miners collude to form so-called \emph{mining pools}, where participants work together towards finding the next block and share rewards based on each miner's contribution and according to some reward distribution scheme. As such, since the appearance of the first mining pools in Bitcoin, the fraction of blocks generated by solo miners has steadily declined and is negligible today. 

The rising competition among mining pools has been shown to motivate adversarial strategies, such as denial-of-service attacks~\cite{johnson2014game,laszka2015bitcoin}.
Other, less detectable, attack techniques such as \emph{block withholding}~\cite{eyal2015miner,luu2015power,courtois2014subversive,rosenfeld2011analysis} and \emph{spy mining}~\cite{sompolinsky2018bitcoin,eyal2014majority} have been observed and studied.
A recent economic model suggests miners will refrain from attacking the underlying blockchain as long as the revenue received from honest behavior exceeds the one-off benefit yielded by attacks~\cite{Budish:2018aa}.

\subsection{Reward Distribution in Mining Pools}\label{sec:background_payout_schemes}

At the time of writing, major pools direct their mining revenue to publicly known addresses, specified in the output of a block's coinbase transaction.
We refer to this address as \textit{reward address}. 
Furthermore, mining pools have been observed to usually reveal their identity by adding human-readable text or \emph{markers} to the coinbase. 
The reward address, often used for prolonged periods by a pool, can thereby be used (i) to directly distribute block rewards to pool members, which can be identified by a set of \textit{miners' addresses}, or (ii) to send the newly minted coins to a \textit{collector address}, which is used to hold larger amounts of BTC so as to perform payments to pool members when necessary. 
We refer to the transaction by which pool members receive their mining reward share as \textit{payout transaction}. 
Such transactions are often characterized by a large number of output addresses, each receiving a reward usually in the order of a few mBTC, as our findings show in Section~\ref{subsec:level2}. 

While the exact structures of payout schemes vary from mining pool to mining pool and also with time, the general principle is the same: mining pool operators distribute the template for the next to-be-generated block to participating miners and require the latter to submit PoW solutions meeting some minimal difficulty.
These ``partial'' solutions are referred to as \emph{shares} and serve as a measure of each miner's contribution. 
Using this information, the mining pool operators distribute the revenue among participating miners, based on some pre-defined scheme. 
An overview of the main classes of reward distribution schemes employed by mining pools, as well as discussions on fairness, is provided in~\cite{rosenfeld2011analysis,schrijvers2016incentive}.

%% file: sections/analysis.tex
\section{Analysis}\label{sec:analysis}

As a starting point of our investigation, in Section~\ref{subsec:level1}, we quantify the overall market share of Bitcoin mining pools by attributing Bitcoin blocks to known mining pools.
Next, in Section~\ref{subsec:level2}, we identify individual miners and investigate revenue streams, obtaining a more accurate picture of the distribution of mining rewards within pools.
Finally, in Section~\ref{subsec:cross-mining}, we investigate the economic relationships between pools and other actors in the Bitcoin ecosystem. 
The analysis results in the rest of this paper are based on Bitcoin blocks 0-\currentBlockheight \xspace(3 Jan. 2009 - 31 Dec. 2018).

\input{sections/analysis_level1}

\input{sections/analysis_level2}

\input{sections/analysis_cross_pool}

%% file: sections/analysis_level1.tex
\subsection{Market Shares of Bitcoin Mining Pools}\label{subsec:level1}


The first step of our analysis consists of attributing mined blocks to mining entities, i.e., pools and individual miners. 
Thereby, we leverage two main information sources: coinbase markers and coinbase transaction output addresses.

We have seen that the \emph{coinbase field} of the coinbase transaction, is usually used by miners to place so-called \emph{coinbase markers}~\cite{judmayer2017merged} --- also called \emph{coinbase tags} or \emph{signatures} --- in order to claim blocks publicly.
This enables mining pool members to monitor their respective pool activity and allows estimating the overall mining power of pools. 
Providing this information is also relevant for miners to publicly show that they support certain forks by setting the respective version bits using a signaling mechanism\footnote{\url{https://github.com/bitcoin/bips/blob/master/bip-0135.mediawiki}}.
As already noted in~\cite{judmayer2017merged}, this information is not cryptographically secured and hence can easily be faked by the miner of a block. 
Therefore, it is reasonable to rely on the \emph{reward address} as a primary data source for block attribution, as modifications to the latter, e.g. to impersonate a different mining entity, result in the loss of the associated mining reward.

In our observation period, 
only $ \percentMultiCBOut\% $ of all blocks exhibited more than one coinbase output address, from which most occurred in the early days of Bitcoin.
We note that coinbase transactions, which have more than one output address, can theoretically contain ordinary payouts performed by the pool and therefore might not be directly mappable to one single entity directly.
For example, P2Pool and Eligius use the coinbase transaction to payout shares of the block reward without an intermediary transaction. 

Although there are several online resources (e.g., \emph{blockchain.info}, \emph{btc.com}, \emph{blocktrail.com}, etc.) which provide aggregated charts regarding the shares miners hold in various cryptocurrencies, their exact methodology of how they attribute blocks to individual miners/pools often remains undisclosed, as for example with \emph{blocktrail.com}. 
If underlying attribution/mapping data is publicly available, it sometimes is outdated, as it is the case for \emph{blockchain.info}, or diverges between services, as it is the case for \emph{btc.com}.

Data provided by \emph{btc.com} was forked from the information provided by \emph{blockchain.info} but tends to use different miner/pool names which makes unification of mapping information a highly manual task. 
To map blocks to mining entities we retrieve mapping information from the following sources and merge them into a single file:
\begin{itemize}
  \item the official \emph{Blockchain.info} Github repository\footnote{\srcBlockchainInfo} that, according to its documentation, is the basis for this visualization \url{https://www.blockchain.com/pools},
  \item the official \emph{BTC.com} Github repository\footnote{\srcBtcCom} that, according to its documentation, is the basis for this visualization \url{https://btc.com/stats/pool}
  \item mappings performed by \emph{Blocktrail.com}\footnote{\url{https://www.blocktrail.com/api}}. However, we do not have precise information about how Blocktrail attributed blocks and this API was closed while working on this paper. Therefore, we don't have Blocktrail mapping information for blocks after $ 514239 $.
  \item manually retrieved coinbase markers from coinbase transactions,
  \item multiple-input cluster information obtained from the GraphSense tool~\cite{github:graphsense}.
\end{itemize}

\begin{figure}
    \centering
    \includegraphics[width=1\columnwidth]{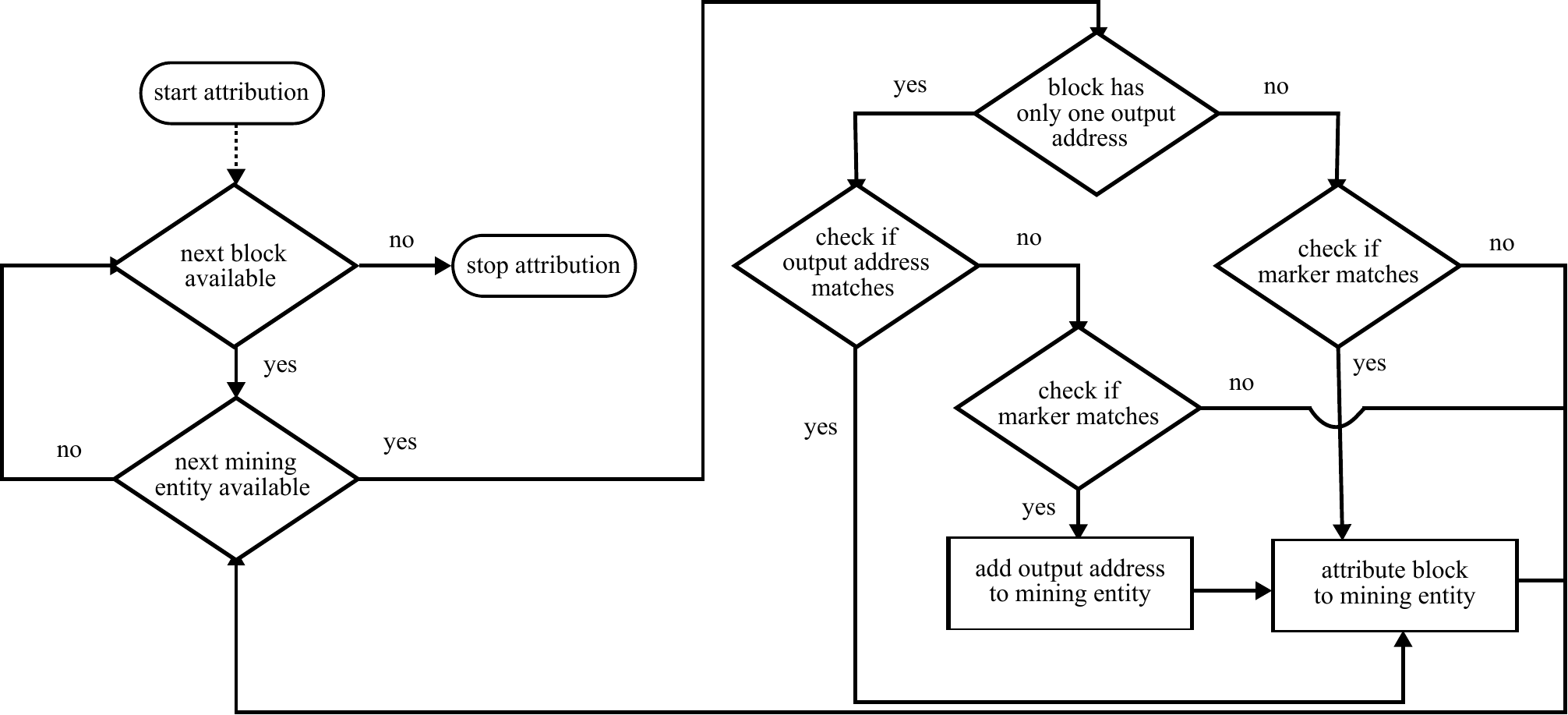}
    \caption{High level flow chart representing our attribution scheme.}
    \label{fig:attribution}
\end{figure}

Given these mappings, our methodology for attributing blocks to mining entities is depicted in Figure~\ref{fig:attribution}. 
First, we unify the names used by different sources to indicate equality of entities in our mapping file. 
This ensures that each mining entity can be uniquely identified, regardless of the source used. 
For each block, we first check whether some mining entity is associated with the coinbase output address.
Recall, we consider this the most reliable source of information for attribution in a block, as providing wrong data here would result in the miner giving away funds. 
If the reward address cannot be identified, e.g in cases where the reward address is not yet attributed to a mining entity or the coinbase transaction has multiple outputs, we proceed to check the coinbase marker. 
If a match is found and there is only one coinbase output address, this address is added to the list of reward addresses of the corresponding mining entity. 
Moreover, the respective mining entity is added to the list of attributions for this block, alongside the source(s) used for the attribution. 

If all sources attribute the same entity to a block, we call this mapping \emph{unique}.
Only in $ \numberConflicts $ out of $ \currentBlockheight $ blocks ($ \percentConflicts\% $), we encountered conflicts, i.e., the block was attributed to two or more mining entities.
Conflicts mostly occurred for blocks that were attributed to \emph{BTC.TOP} and \emph{CANOE} at the same time, as well as Waterhole and BTC.com, or Tang Pool and Bixin.
Despite being just a small fraction of blocks, it shows that publicly available resources for Bitcoin analysis are not always precise and accurate. 
Sometimes even the information coming from a single source is inconsistent and would attribute a single block to different miners.
For example, only using the pool information from \emph{Blockchain.info} blocks $ 482059 $ and $ 482221 $ are attributed to \emph{Waterhole} (based on address) as well as \emph{BTC.com} (based on coinbase).
Other examples can be found in BTC pool information, which is used in our attribution scheme: block $ 524045 $, for instance, is attributed to \emph{BitcoinRussia} and \emph{Bitcoin-Ukraine}; $ 16 $ other blocks ($476422, 478113, 482242, 482614, 483576, \ldots$) are attributed to \emph{CANOE} and \emph{BTC.TOP}.

Overall, with our method, we attributed more blocks than any other source considered alone. 
As an example, if we consider blocks from \attrStart~to \attrEnd~we attributed \attr~blocks, which is \attrMoreBlockchainInfo~($\sim$ \attrMoreBlockchainInfoBTC~BTC) more than it would be possible by only using pool information from \emph{blockchain.info}.
A comparison of different attribution data sources is shown in Table~\ref{tbl:attr}, while the conflicts in our final attributed dataset are distributed as shown in Table~\ref{tbl:conflicts}.

\begin{table}
  \centering
  \caption{Comparison of different attribution data sources for miners and pools from block 0 to $ \currentBlockheight $. 
  The column \emph{direct use} 
  refers to using the provided mapping information directly on every block without adding newly identified addresses by markers.
  The column \emph{our attribution} refers to the procedure described in Figure \ref{fig:attribution}. For \emph{blocktrail.com} we only have a list of blocks up to $ 514,239 $ with the respective associated attribution, therefore there cannot be any conflicts. \emph{Combined} contains all previously mentioned sources as well as GraphSense and manually identified coinbase markers.} \label{tbl:attr}
  \begin{tabular}{@{}lllll@{}}
    \toprule
    Source          & Direct use & \begin{tabular}[c]{@{}l@{}}Direct use \\ conflicts\end{tabular} & Our attribution & \begin{tabular}[c]{@{}l@{}}Our attribution \\ conflicts\end{tabular} \\
    \midrule
    blockchain.info & 334,416    & 2                                                               & 339,597         & 5                                                                    \\
    btc.com         & 337,629    & 3                                                               & 342,619         & 22                                                                   \\
    blocktrail.com  & 324,720    & -                                                               & -               & -                                                                    \\
    combined        & -          & -                                                               & 375,381         & 684                                                                  \\
    \bottomrule
  \end{tabular}
  \vspace{0.2cm}  
\end{table}

\begin{table}
  \centering
  \caption{Conflicts in final attribution. The last conflict between F2Pool and BTCC Pool is probably due to a misattribution by \emph{blocktrail.com} as all other sources attribute the respective block to F2Pool.}\label{tbl:conflicts}
  \begin{tabular}{@{}llll@{}}
  \toprule
  Miner 1       & Miner 2         & Number of conflicts & Example blocks         \\
  \midrule
  BTC.TOP       & CANOE           & 338                 & 516210, 516275, \ldots   \\
  Bixin         & TangPool        & 142                 & 339210, 339284, \ldots    \\
  BTC.com       & Waterhole       & 113                 & 478230, 478328, \ldots    \\
  BTC.TOP       & WAYI.CN         & 81                  & 509073, 509100, \ldots    \\
  ViaBTC        & Okminer         & 5                   & 510279, 523217,        \\
  Yourbtc       & OzCoin          & 3                   & 159846, 159929, 159964 \\
  BitcoinRussia & Bitcoin-Ukraine & 1                   & 524045                 \\
  F2Pool        & BTCC Pool       & 1                   & 482886                 \\
  \bottomrule
  \end{tabular}
  \vspace{0.2cm}
\end{table}



\subsubsection{Evolution of mining pool market shares.}

\begin{figure}
    \hspace*{-1.7cm}\includegraphics[width=1.2\textwidth]{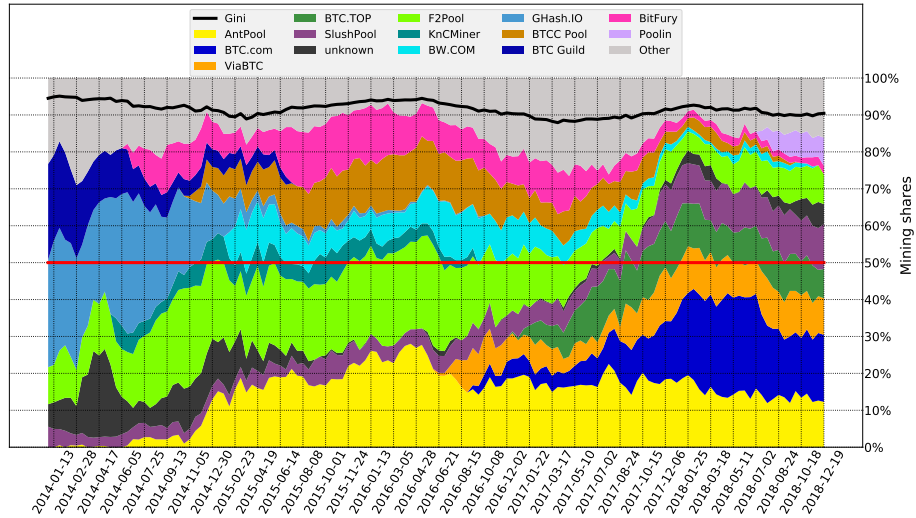}
    \caption{Evolution of mining pool market shares in Bitcoin between \stackplotStart~and \stackplotEnd. The red line indicates the $ 50\% $ security threshold, while the black line is the Gini coefficient as a measure of market share distribution.}\label{fig:stack_miners}

\end{figure}

Having attributed blocks to mining pools, we can now analyze how their shares have evolved over time.
Figure~\ref{fig:stack_miners} shows the evolution of mining shares between \stackplotStart~and \stackplotEnd, aggregated in bins spanning $ 2,016 $ blocks, which corresponds to Bitcoin's difficulty adjustment period.
The gray region represents small known pools or miners for which the sum of all its percentages is below $ 4\% $. 
It also indicates the $ 50\% $ mining power threshold (red line) and the Gini coefficient (black line) as a measure of market share distribution (between 0 and 1). 
The higher the Gini value, the stronger the inequality among market's participants.
The evolution of the Gini coefficient shows peaks of mining power centralization around June 2014, April 2016 and January 2018, almost in a cyclical fashion (roughly a 22-month period).

Regarding the evolution of mining pools' market shares, we can observe that the distribution has changed over time.
Pools dominating mining now (BTC.com, ViaBTC and BTC.TOP) did not exist in early 2016 and, vice versa, most of the largest pools in 2016 (e.g., BTCC, Bitfury and BW Pool) are now much smaller players. 
We can also observe that, from January until mid-2018, three pools combined held more than 50\% of the overall mining power.
In particular, the first two (BTC.com and AntPool) are owned by Bitmain, the leader among mining hardware manufacturers. 
Another interesting observation is that the number of \emph{unknown} blocks which could not be attributed to a mining entity has increased lately to a level last observed 2015. 

%% file: sections/analysis_level2.tex
\subsection{Mining Reward Distribution}\label{subsec:level2}


In the previous section, we saw how mining shares are distributed among pools and how they evolved over time. Before we can further investigate revenue streams within and across mining pools, we need to understand how pool operators split the block reward among individual miners and identify as many payout transactions as possible. From now on, we will focus our analysis on three pools (BTC.com, AntPool and ViaBTC) that held the majority of mining power at the beginning of 2018. Since payout distribution schemes also change over time, we further limit our study to a four-week observation period ranging from block 510,000 (2018-02-19) to 514,032 (2018-03-18) when each major pool almost always followed a distinctive and stable payout pattern (see Figure~\ref{fig:patterns}).
This allows us to identify payout transactions while reducing the number of false positive transactions that do not represent payments to individual miners. 
To verify that the identified payout transactions are indeed within reasonable bounds, we compare the amount of BTC paid out via these transactions with the amount of BTC received by the pool in the same period, as described in detail in Section~\ref{subsec:cross-mining}. 
We identify transactions from mining pools to individual miners using the following, pool-specific heuristics:

\begin{figure}
	\centering
	\qquad
		\subfloat[1][BTC.com payout pattern.]{\label{fig:btccom_flow}
		\includegraphics[width=0.6\textwidth]{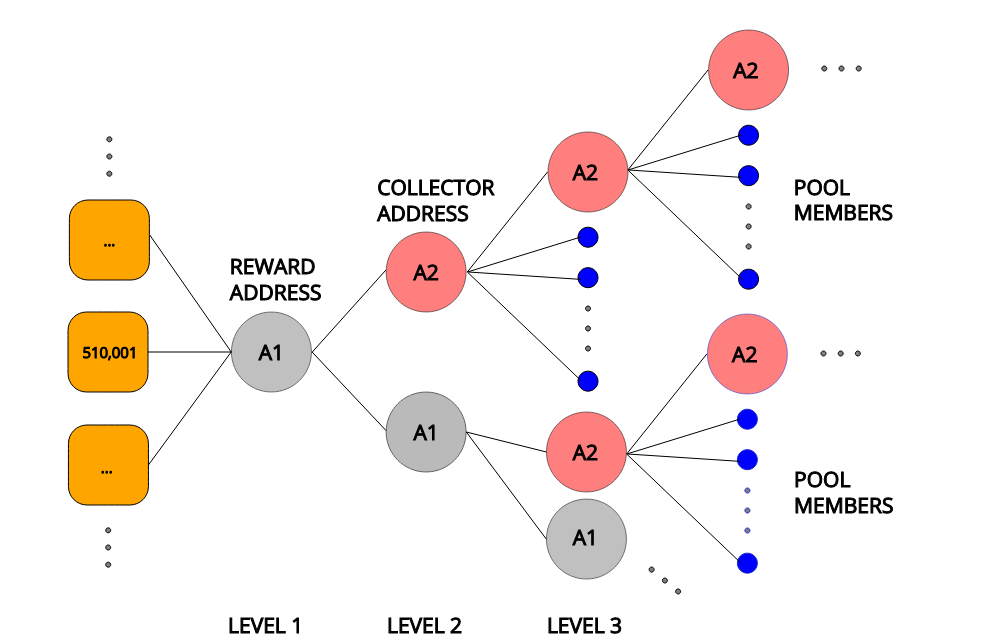}}
		\subfloat[2][ViaBTC payout pattern.]{\label{fig:viabtc_flow}
		\includegraphics[width=0.4\textwidth]{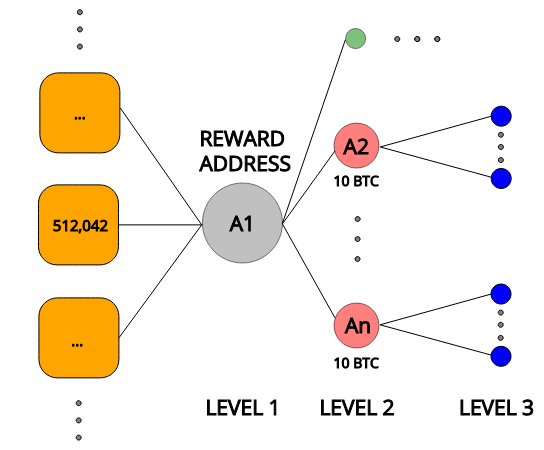}}

		\subfloat[3][AntPool payout pattern.]{\label{fig:antpool_flow}
		\includegraphics[width=0.65\textwidth]{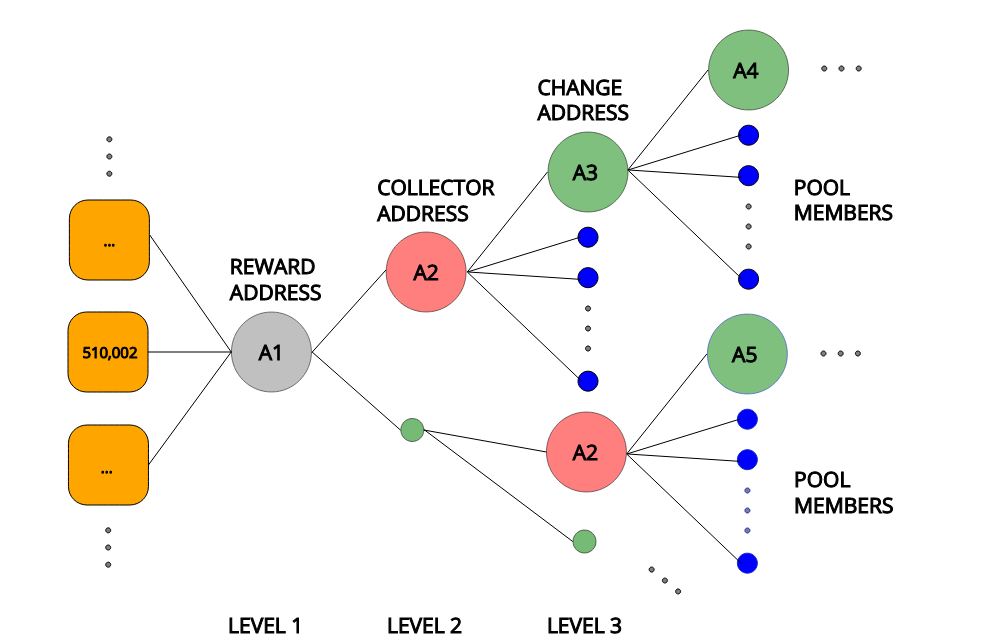}}
	\caption{Payout patterns observed in the time period between block 510,000 and 514,032. In gray: reward addresses, in red: addresses performing payout transactions, in blue: pool members, in green: change addresses. Rounded squares are coinbases of blocks mined by the pool. The size of the nodes indicates the differences in received BTC per transaction per address. In BTC.com (Figure~\ref{fig:btccom_flow}) payout transactions are performed by one single collector address. In ViaBTC (Figure~\ref{fig:viabtc_flow}), payout transactions are performed by a dozen of addresses (always changing), receiving 10 BTC each from one single reward address. In AntPool (Figure~\ref{fig:antpool_flow}), similarly to BTC.com, we have a payout chain that originates from always the same collector address, but continues with other change addresses used only once as input in payout transactions.}\label{fig:patterns}\end{figure}

\begin{itemize}

	\item \textbf{BTC.com} received block rewards always in the same reward address denoted as $A_1$ in Figure~\ref{fig:btccom_flow} and used one collector address, denoted as $A_2$, for distributing mining rewards to pool members. That address was also used as change address in payout transactions and the payments continue in a chain-like fashion. Supposing that only this address was used to transfer funds to pool members, we selected all its payout transactions within the examined period, each having a large number of outputs (in the order of $10^3$) with relatively low amounts of revenues associated (few mBTC).

	\item \textbf{AntPool} also collected most of the mined coins using a single address. As in BTC.com, they have always been sent to one collector address responsible for payments ($A_2$ in Figure~\ref{fig:antpool_flow}), but, differently from BTC.com, the chain of payments continues with a series of change addresses that are never reused. A peculiar aspect of these payout transactions is that a significant percentage of them\footnote{The exact percentage and more details are provided in the Section~\ref{subsec:cross-mining}.} has 101 output addresses (change address included). 
	Because of that, in order to identify the change address in a payout transaction, we investigate all output addresses that spent the received amount through another payout transaction with 101 output addresses. In the time period considered, it turns out that the change addresses were always the output with the largest sum of BTC. 
	By following this series of change addresses and payout transactions, we identified several other addresses as being individual miners, i.e., members of the pool.

	\item \textbf{ViaBTC} followed a payout pattern similar to the one represented in Figure~\ref{fig:viabtc_flow}, where a dozen of addresses always received a sum of exactly 10 BTC, which was then spent in payout transactions, again with a few hundred outputs and small sums of BTC. Since the dozen of addresses was not fixed, we investigate all transactions that received 10 BTC from the ViaBTC reward address. We must report that, in this case, we could not consistently distinguish the pool's change address from the members' addresses.

\end{itemize}


\subsubsection{Identification of individual miners.} 

Having identified payout transactions, we can now extract Bitcoin addresses belonging to individual miners. Furthermore, we can partition these addresses into maximal subsets (clusters) that are likely to be controlled by the same actor using the well-known~\cite{reid2011anonymity,ron2013quantitative} and efficient~\cite{Harrigan:2016aa} multiple-input clustering heuristics.
The underlying intuition is that if two addresses (i.e.: A and B) are used as inputs in the same transaction while one of these addresses along with another address (i.e: B and C) are used as inputs in another transaction, then the three addresses (A, B and C) must somehow be controlled by the same actor~\cite{meiklejohn2013fistful}, who conducted both transactions and therefore possesses the private keys corresponding to all three addresses. 
This heuristic fails when CoinJoin transactions~\footnote{\url{https://en.bitcoin.it/wiki/CoinJoin}} are taken into account because they combine payments from different spenders that do not necessarily represent one single entity. Aware of this problem, we filtered those transactions before applying the multiple input heuristics.

Table \ref{miners_stats} provides summary statistics for each investigated mining pool: the number of blocks mined by each pool ($N_{B}$), the number of identified payout transactions ($N_{TX}$), the number of addresses belonging to individual miners ($N_{A}$) and the number of identified clusters ($N_{C}$) within each pool. In order to estimate the real-world coverage of our dataset, we compared the mining reward ($BTC_{M}$) associated to mined blocks with the payouts associated to individual miners' addresses ($BTC_{P}$). This shows that we were---within our observation period
and provided that payouts happen regularly---able to identify 92\% of the individual miners in BTC.com, 30\% in AntPool and 75\% in ViaBTC. 
We hypothesize that the low percentage obtained for AntPool lies on the relatively strict filter criterion we applied on its payout pattern (transactions with exactly 101 outputs). 
We also investigated how often individual miners reused addresses within a pool and found that the median address reuse $\mu$ is much higher in BTC.com than in AntPool and ViaBTC.
When we normalize the median address reuse by the number of identified payout addresses $\dfrac{\mu}{N_A}$, we see that AntPool is the outlier, which could mean that its members are more careful about privacy or are changing their payout addresses at a faster rate compared to the other pools.
The fact that the number of blocks mined is greater than the number of payments we detected can be due to two reasons: pool managers distribute the mining rewards not at every mined block, but within a longer time period, combining payments to minimize the transaction fees and, as already noted above, because we didn't manage to find all payout transactions performed by pools.

\setlength{\tabcolsep}{4pt}
\begin{table}
	\centering
	\caption{Statistics of retrieved data within the observed period. $N_{B}$: number of blocks mined by the pool, $N_{TX}$: number of identified payout transactions, $N_{A}$: number of identified members' addresses, $N_{C}$: number of identified clusters, $BTC_{M}$: BTC mined by the pool, $BTC_{P}$: BTC paid to pool members (addresses), $\mu$: median value of address reuse.}\label{miners_stats}
	\begin{tabular}{@{}lrrrrrrrrrr@{}}
	\toprule
	Pool Name & 
	$N_{B}$ & 
	$N_{TX}$ & 
	$N_{A}$ & 
	$N_{C}$ & 
	$BTC_{M}$ &
	$BTC_{P}$ &
	$\dfrac{BTC_{P}}{BTC_{M}}$ & 
	$\mu$ &
	$\dfrac{\mu}{N_A}$ \\
	\midrule
	BTC.com 	&1,020	&225	&20,444	&8,900	&13,059	&12,057	&92\%	&20 &9.8 $\times10^{-4}$ \\
	AntPool 	&617	&408	&14,166	&5,082	&7,887	&2,333	&30\%	&2 &1.4 $\times10^{-4}$\\
	ViaBTC 		&457	&104	&7,171	&3,121	&5,841	&4,284	&75\%	&5 &7.0 $\times10^{-4}$\\
	\bottomrule
	\end{tabular}	
\end{table}

\subsubsection{Centralization of mining shares within pools.}

\begin{figure}
	\centering
	\includegraphics[width=0.8\textwidth]{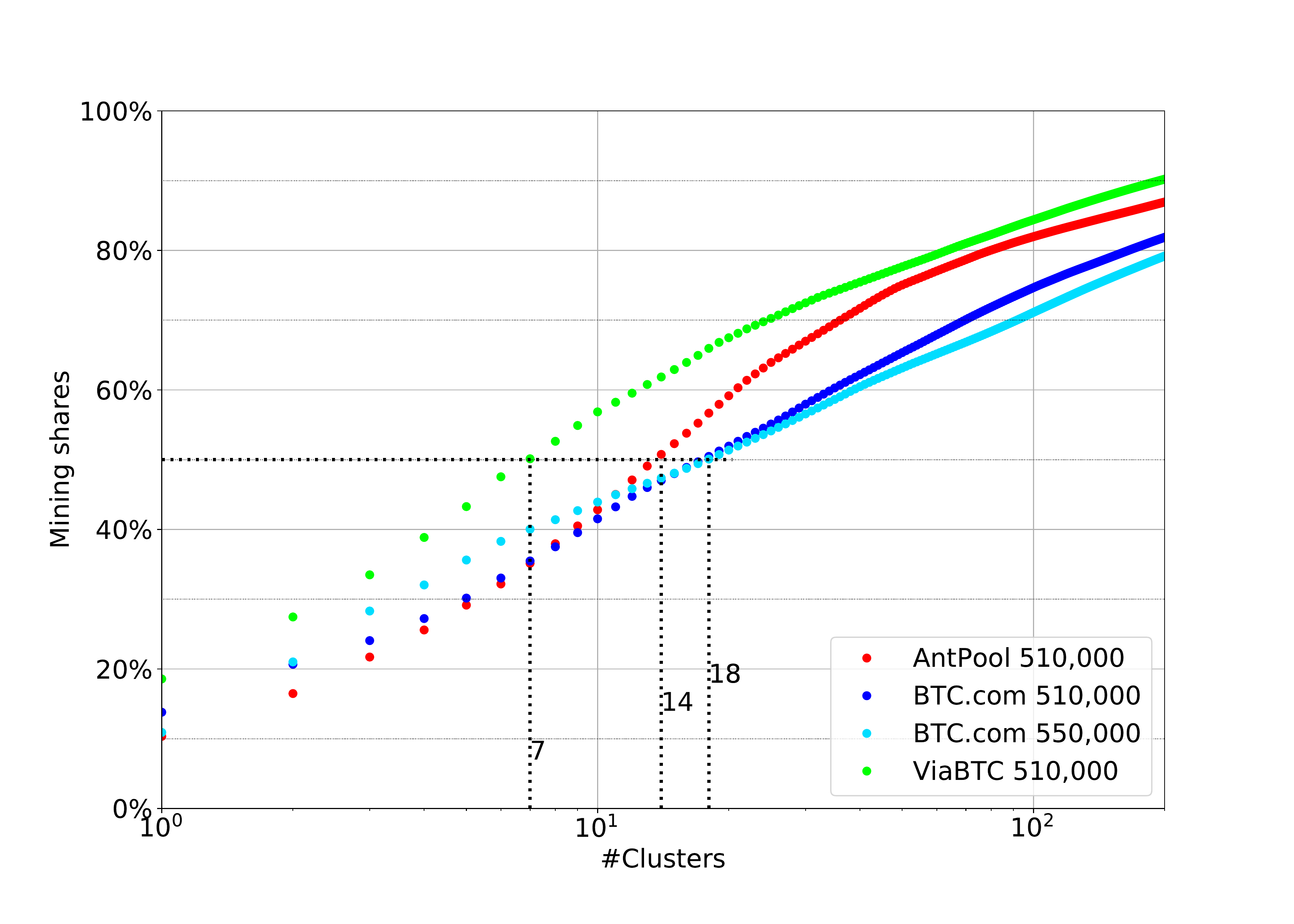}
	\caption{Cumulative sum of mining shares over clusters (actors) for each pool (log-scale). Black-dotted lines highlight the number of clusters controlling 50\% of each pool.} \label{fig:pool_centralization}
\end{figure}

Previously, in Section~\ref{subsec:level1}, we saw that the mining shares are centralized among a relatively small number of pools. However, little is known about the centralization of mining shares inside each pool. We gain insight into pool centralization by looking at the distribution of a pool's mining shares to identified clusters ($N_{C}$), which represent actors within each pool. Figure~\ref{fig:pool_centralization} shows the cumulative distribution of mining shares among members (clusters) for each pool.

In order to investigate how the internal mining distribution changed over time, we expand our dataset (blocks 510,000 to 514,032) by payout transactions for BTC.com from block 550,000 (2018-11-14) to 554,032 (2018-12-16). We chose BTC.com because its payout transactions can easily be identified, as discussed before. In Figure~\ref{fig:pool_centralization} the blue and light-blue lines show the cumulative sum of BTC.com for both periods and it can be observed that the distribution of mining shares within that pool remained relatively stable over time.
 
Although our dataset covers just a fraction of Bitcoin's overall mining activity, we notice that 50\% of the mining power in ViaBTC is controlled by 7 clusters, compared to 20 clusters for BTC.com and 15 clusters for AntPool. Despite this, if we compute the Gini coefficient on these shares we get 0.945 for BTC.com, 0.942 for ViaBTC and 0.938 for AntPool, which indicates that the distribution of mining shares is highly centralized within all investigated pools. We note and discuss later that clusters do not necessarily represent individuals but also larger actors such as exchanges or wallet providers.

%% file: sections/analysis_cross_pool.tex
\subsection{Miners in the Bitcoin Ecosystem}\label{subsec:cross-mining}

Having identified individual miners within pools by their Bitcoin addresses and cluster affiliation, we can now turn our focus on individual miners participating in several pools and lay out their economic relationships to other actors in the Bitcoin ecosystem within our observation period. 

\subsubsection{Cross-Pool Mining}

If a single Bitcoin address receives payouts from several mining pools, we can assume that the individual miner holding that address conducts cross-pool mining. Table~\ref{table:crosspool_addresses} shows the number of addresses involved in cross-pool mining for each pair of pools, their respective clusters, as well as the total amount of BTC received by these entities from the pools. We notice that the BTC.com-AntPool pair is the one with the highest overlaps (addresses in common, clusters in common and BTC received), which might be a result of the common mining pool ownership as discussed in Section~\ref{subsec:level1}. However, it must be noted that these figures only represent the fraction of addresses that were reused across clusters; individual miners creating separate addresses for each of their mining activities are immune to multiple-input clustering and therefore not represented in this table.

\begin{table}
\centering
\caption{Cross-pool mining at address level in the time period between block 510,000 and 514,032 ($\sim$ 4 weeks), including how much BTC from each pool has been received by those common addresses.} \label{table:crosspool_addresses}
\begin{tabular}{@{}llllll@{}}
\toprule
Pool 1  & Pool 2  & \begin{tabular}[c]{@{}l@{}}Addresses \\ in common\end{tabular} & \begin{tabular}[c]{@{}l@{}}Clusters \\ in common\end{tabular} & \begin{tabular}[c]{@{}l@{}}BTC from\\ Pool 1\end{tabular} & \begin{tabular}[c]{@{}l@{}}BTC from\\ Pool 2\end{tabular} \\ \midrule
BTC.com & AntPool & 537 (1.58\%) & 434 (3.2\%)        & 664.3 (5.5\%)                                               & 176.8 (7.6\%)                                                \\
AntPool & ViaBTC  & 115 (0.54\%) & 196  (2.4\%)      & 11.1  (0.47\%)                                               & 102.6 (2.4\%)                                                \\
ViaBTC  & BTC.com & 250 (0.91\%)  & 267  (2.3\%)     & 175.4 (4.1\%)                                               & 174.1 (1.4\%)                                               \\ \bottomrule
\end{tabular}
\end{table}

Next, we uncover the actors behind the clusters reported in Table~\ref{table:crosspool_addresses} using publicly available tags from blockchain.info and walletexplorer.com. 
In Table~\ref{table:crosspool_entities} we report the main actors that have been receiving shares of block rewards from the three pools analyzed. 
For each entity-pool pair, the table shows the amount of BTC received by the entity from the pool, its share within the pool and the number of addresses associated with the cluster representing that actor. 
The majority of actors is \emph{Unknown}, which means that we could not find tags attributing the associated addresses to actors.
Excluding this last set which together accounts for more than 65\% of the shares in each pool, we see that the mining rewards are distributed to cryptocurrency exchanges (E) and wallet providers (W). The top exchanges listed in Table~\ref{table:crosspool_entities} (Bixin, Huobi.com) hold, in combination, relatively strong mining shares within each pool (BTC.com: 20,51\%, AntPool: 16,43\%, ViaBTC: 24.02\%) and can, therefore, be regarded as major forces pushing the centralization of mining shares within pools, as reported before in Section~\ref{subsec:level2}. However, it must be noted that exchanges typically do not participate in mining activities themselves but host wallets of individual miners. Nevertheless, we point out that exchanges and wallet providers are usually operated by a single physical or legal entity and the ownership of assets is often unclear unless users withdraw cryptocurrency units into self-controlled hot or cold wallets~\cite{anderson2018redux}.

Furthermore, we can also observe a geographical co-location of mining pool operators and payout services: the same top exchanges have (or had\footnote{\url{https://www.forbes.com/sites/kenrapoza/2017/11/02/cryptocurrency-exchanges-officially-dead-in-china/\#48116a6a2a83}}) strong ties with China, which is where the three observed mining pool operators are located. 

\setlength{\tabcolsep}{2pt}
\begin{table}
\caption{Cross-pool mining at a cluster level in the time period between block 510,000 and 514,032 ($\sim$ 4 weeks). For each unknown entity or known actor, it is shown the amount of BTC received by each pool, its share in the pool, the number of addresses linked to it and the type of service it offers (W: wallet provider, E: cryptocurrency exchange service, P: known mining pool, M: unknown mining entity).}\label{table:crosspool_entities}
\scalebox{0.73}{
\begin{tabular}{@{}llrrrrrrrrrr@{}}
\toprule
              &  & \multicolumn{3}{c}{BTC.com}   & \multicolumn{3}{c}{AntPool}   & \multicolumn{3}{c}{ViaBTC}    &  \\ \cmidrule(lr){3-5} \cmidrule(lr){6-8} \cmidrule(lr){9-11}
Entity/Actor             & Service        & BTC     & \%BTC & \#Addr. & BTC     & \%BTC & \#Addr. & BTC     & \%BTC & \#Addr. & Total BTC \\ \midrule
Unknown             & ?       & 8930.39 & 74.07 & 13286       & 1682.25 & 72.09 & 8888        & 2877.02 & 67.17 & 4845        & 13489.67  \\ 
Bixin               & W+E+P   & 1663.75 & 13.80 & 1061        & 241.28  & 10.34 & 546         & 795.36  & 18.57 & 476         & 2700.39   \\ 
Huobi.com         & E       & 808.64  & 6.71  & 964         & 142.04  & 6.09  & 759         & 225.50  & 5.27  & 322         & 1176.19   \\ 
Bittrex.com         & E       & 83.71   & 0.69  & 348         & 29.56   & 1.27  & 251         & 43.36   & 1.01  & 177         & 156.63    \\ 
Xapo.com            & W       & 26.96   & 0.22  & 94          & 70.75   & 3.03  & 64          & 5.79    & 0.14  & 33          & 103.50    \\ 
Poloniex.com        & E       & 42.65   & 0.35  & 381         & 11.52   & 0.49  & 268         & 19.97   & 0.47  & 139         & 74.15     \\ 
Luno.com            & W+E     & 36.59   & 0.30  & 258         & 4.06    & 0.17  & 104         & 4.39    & 0.10  & 60          & 45.04     \\ 
Bitstamp.net        & E       & 8.94    & 0.07  & 57          & 3.55    & 0.15  & 38          & 3.91    & 0.09  & 22          & 16.39     \\ 
Cryptonator.com     & W+E     & 5.75    & 0.05  & 80          & 0.70    & 0.03  & 41          & 2.70    & 0.06  & 33          & 9.15      \\ 
BitoEX.com          & W       & 5.09    & 0.04  & 23          & 1.12    & 0.05  & 35          & 2.19    & 0.05  & 4           & 8.39      \\ 
CoinHako.com        & W+E     & 3.59    & 0.03  & 4           & 0.29    & 0.01  & 3           & 0.24    & 0.01  & 2           & 4.12      \\ 
Bitcoin.de          & E       & 1.86    & 0.02  & 26          & 0.76    & 0.03  & 13          & 0.58    & 0.01  & 7           & 3.19      \\ \bottomrule
\end{tabular}
}
\end{table}	

\subsubsection{Economic relationships in the Bitcoin ecosystem}

\begin{figure}
    \centering
    \includegraphics[width=0.7\columnwidth]{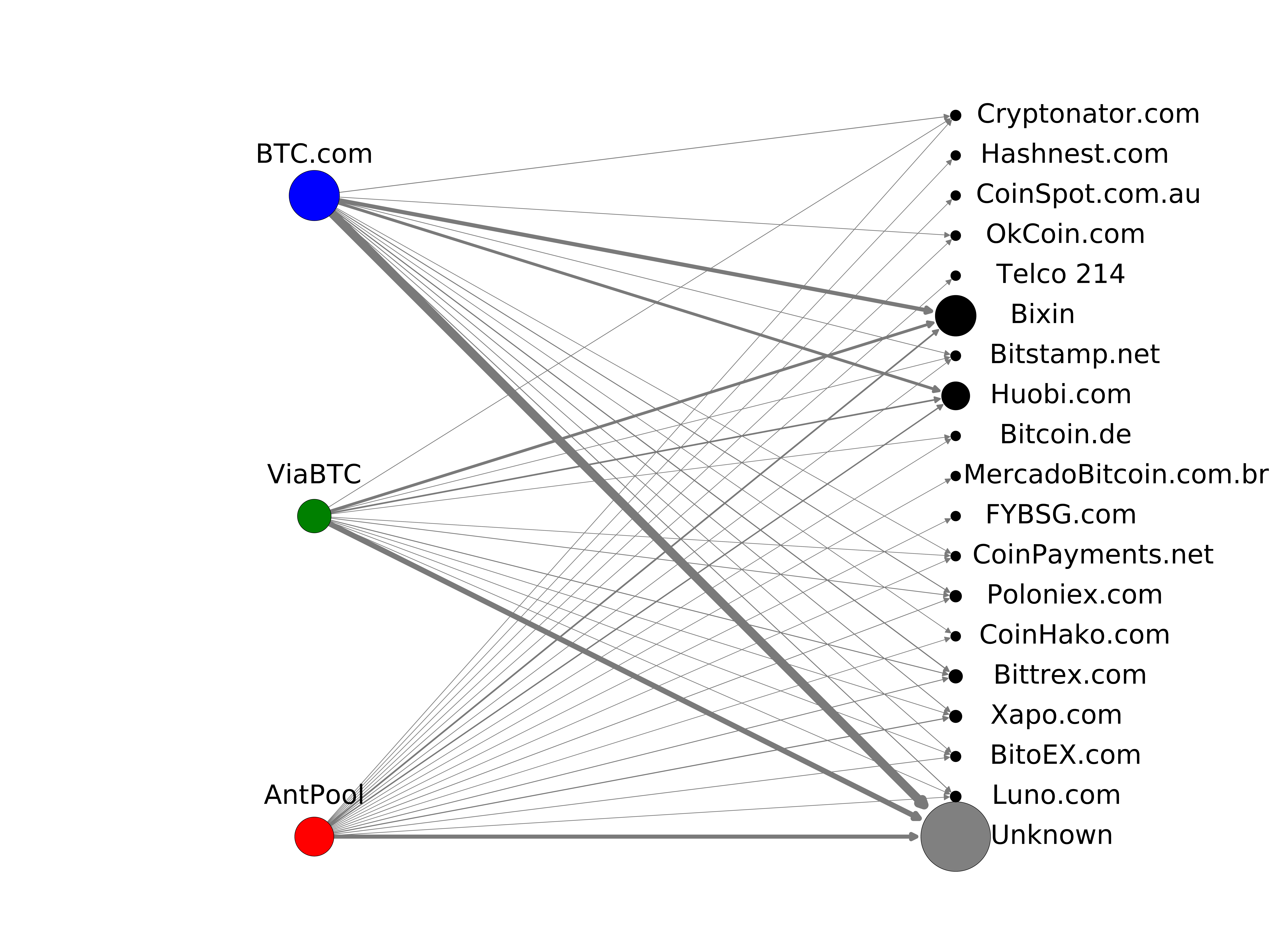}
    \caption{Flow of mining rewards from mining pools to their members. The strength of the arcs is scaled by payment volume, while the node size depends on the total amount of received (mined) BTC. In black: wallet services and exchanges, in gray: unknown entities. This plot covers the top 400 clusters from each mining pool sorted by received BTC. \textit{Unknown} entities (1118) were combined into one node.}
    \label{fig:payment_graph}
\end{figure}

Having identified and partly de-anonymized the actors that were mining with the three analyzed pools within our observation period, we can now illustrate the economic relationships among mining pools and other actors in the Bitcoin ecosystem. Figure~\ref{fig:payment_graph} shows the flow of mining rewards from mining pools to the clusters representing actors within pools. We selected the first 400 clusters\footnote{Sorted by received BTC. This number covers at least 85\% of the mining shares for each pool.} from each pool and grouped together clusters representing unknown entities in one node (1,118 in total). 
From our analysis, it is clear that the vast majority of mined coins go to unknown entities that we grouped into one \textit{Unknown} entry in Table~\ref{table:crosspool_entities}.

\subsubsection{Inspecting the Unknown}

We now focus on the 10 largest entities within the \emph{Unknown} by inspecting basic statistical properties of the underlying address clusters. In Table~\ref{table:crosspool_unknown_entities}, we report for each cluster its internal ID (assigned by GraphSense), the amount of BTC received by each cluster from each pool, the total revenues from mining, as well as the total amount of BTC received (as of April, 23rd 2018).
When further inspecting the total number of addresses in those clusters we can observe that all of them consist of more than 30,000 addresses and one of them (cluster 324067473 with 11,534,706 addresses) is a so-called super-cluster~\cite{Harrigan:2016aa}. 9 clusters have been receiving a relatively large number of transactions for more than a year, 2 clusters (cluster 327539880 and 324067473) for more than 4 years. While not being verifiable without having attribution data, those statistics suggest that the ten largest unknown mining clusters also represent untagged exchange services or wallet providers.


\begin{table}[]
\centering
\caption{Cross-pool mining of the ten largest unknown mining clusters sorted by total amount of BTC received by the three pools in the time period between block 510,000 and 514,032 ($\sim$ 4 weeks).}\label{table:crosspool_unknown_entities}

\scalebox{0.9}{
\begin{tabular}{@{}rrrrrrrrr@{}}
\toprule
           & \multicolumn{2}{c}{BTC.com}                         & \multicolumn{2}{c}{AntPool}                         & \multicolumn{2}{c}{ViaBTC}                          &                      &             \\ \cmidrule(lr){2-3} \cmidrule(lr){4-5} \cmidrule(lr){6-7}
Cluster ID & \multicolumn{1}{c}{BTC} & \multicolumn{1}{c}{\%BTC} & \multicolumn{1}{c}{BTC} & \multicolumn{1}{c}{\%BTC} & \multicolumn{1}{c}{BTC} & \multicolumn{1}{c}{\%BTC} & \multicolumn{1}{c}{\begin{tabular}[c]{@{}c@{}}Mined\\BTC\end{tabular}} & \multicolumn{1}{c}{\begin{tabular}[c]{@{}c@{}}Total BTC\\Received\end{tabular}}     \\ \midrule
327539880  & 409.34                  & 3.40                      & 122.10                  & 5.23                      & 258.55                  & 6.04                      & 789.99               & 521,939  \\
324067473  & 295.02                  & 2.45                      & 90.44                   & 3.88                      & 189.15                  & 4.42                      & 574.61               & 3,756,583  \\
350822682  & 244.77                  & 2.03                      & 9.29                    & 0.40                      & 182.92                  & 4.27                      & 436.98               & 110,566 \\
350824718  & 244.67                  & 2.03                      & 65.65                   & 2.81                      & 46.20                   & 1.08                      & 356.52               & 112,680 \\
333653856  & 153.02                  & 1.27                      & 54.02                   & 2.31                      & 83.60                   & 1.95                      & 290.63               & 130,680   \\
372448840  & 181.10                  & 1.50                      & 33.64                   & 1.44                      & 55.73                   & 1.30                      & 270.48               & 882,713  \\
234254928  & 93.31                   & 0.77                      & 27.18                   & 1.16                      & 58.68                   & 1.37                      & 179.17               & 905,101   \\
249123673  & 15.63                   & 0.13                      & 0.40                    & 0.02                      & 107.23                  & 2.50                      & 123.26               & 6,812,938 \\
349962609  & 8.67                    & 0.07                      & 39.01                   & 1.67                      & 19.74                   & 0.46                      & 67.41                & 1,173,892 \\
311503667  & 38.94                   & 0.32                      & 7.47                    & 0.32                      & 7.77                    & 0.18                      & 54.18                & 486,338 \\
\bottomrule
\end{tabular}
}
\end{table}


%% file: sections/discussion.tex
\section{Discussion}\label{sec:discussion}

We believe that the empirical analysis presented in this paper led to novel insights into the structure and behavior of Bitcoin mining pools. 
Our longitudinal analysis of mining pool market shares, 
which is based on a simple yet effective attribution scheme, outlines the need for open attribution data 
and agreed upon reproducible methodologies of how this attribution data is applied. 
Moreover, it highlights conflicts and gray spots in block attributions not visible in aggregated pie charts and confirms centralization tendencies in Bitcoin mining, as pointed out by previous research. 
It also confirms concerns targeting the recent domination of three to four mining pools, which could surpass the 50\% security threshold when combining their mining power. 

Additionally, we were able to trace payout transactions within a one-month observation period, found addresses belonging to individual miners and could group them into clusters representing actors in the Bitcoin ecosystem. We showed that the distribution of shares within analyzed mining pools is also highly centralized and that the majority of mining rewards distributed by a pool is received by a relatively small set of actors. We also saw that individual miners conduct cross-pool mining and
that geographically co-located exchange services and wallet services hold large shares within pools and push towards centralization within mining pools. Plus, our dataset is openly available and our method is reproducible and could be extended over additional mining pools, longer observation periods and other cryptocurrencies.

We are well aware that our approach has a number of limitations. First, we focus our in-depth analysis of mining pools on a restricted set of pools (the top three pools) and an observation period of one month. For AntPool in particular, we selected only payout transactions following the pattern 101 outputs, even though, by manual investigation we noticed that the chain of payments sometimes continued with a different amount of outputs. Improvements in this aspect are possible and are part of our future works.
Other, non-investigated smaller pools could be less centralized and follow other payout patterns.
We address this limitation, by binding our results to specific, in our opinion systemically relevant, mining pools.
This ensures, that we are not claiming that we assess the entire set of mining pools in Bitcoin.
Second, we know that the multiple-input heuristics we used for clustering addresses could lead to false positives (unrelated addresses are joined together) when transactions are tunneled through Mixing services (c.f.,~\cite{MoeserB2016JoinMeOnAMarket}) and false negatives (mining addresses do not belong to any cluster) when individual miners create new addresses for each single mining activity. However, we are confident of having avoided false positives by applying strict filtering criteria on mining pool payout patterns and ignoring CoinJoin transactions. Third, our approach is limited by the extent and quality of the attribution data (tags) available. Without this information, address clusters remain anonymous and inferences about their real-world nature are impossible. Nevertheless, we believe that such data will increasingly become available in the near future with the growing popularity of cryptocurrency analytics tools.

Overall, our paper strengthens a line of recent research and community discussions that suggest skepticism and scrutiny on the decentralization of control in cryptocurrencies, which is often considered being the key feature distinguishing them from fiat currencies. 
To plausibly uphold this claim, mining pools as well as big players in the ecosystem like exchanges, have to find the sweet spot between acting more transparently to encourage public auditability and the privacy demands of their users. 


%% file: sections/conclusions.tex
\section{Conclusions}\label{sec:conclusions}


We present an empirical analysis of the distribution of mining shares within and across mining pools.
Our investigation on the longitudinal evolution of mining pools confirms centralization among a relatively small number of mining pools, three to four controlling more than 50\% of the hash rate. 
Further inspection of the three largest mining pools has shown centralization tendencies also within those pools, where in each pool less than 18 pools members receive more than 50\% of the identified pool payouts.
Examination of payments between those mining pools and actors representing individual miners has revealed cross-pool mining activity (both at address and cluster level), economic relationships between operators of geographically co-located mining pools, exchange services and wallet providers. Overall, our research supports previous findings and scrutinizes the decentralization property of cryptocurrencies.


%% file: sections/acknowledgments.tex
\section{Acknowledgments}\label{sec:acknowledgments}

This work was funded by the Austrian Research Promotion Agency (FFG) through the projects VIRTCRIME, SESC and PR4DLT (Project IDs: 860672, 858561 and 864738), the competence center SBA-K1 (funded by COMET) and Blockchain.com.